\documentclass[a4paper,11pt]{article}
\usepackage{pos}
\usepackage{hyperref}

\title{Recent heavy-flavour measurements from ALICE}

\author[]{Jonghan Park$^{a,*}$ for the ALICE Collaboration}

\affiliation[a]{University of Tsukuba,\\
  1 Chome-1-1 Tennodai, Tsukuba, Ibaraki, Japan}

\emailAdd{jonghan@cern.ch}

\abstract{
Studying heavy-flavour mesons and baryons in hadronic collisions provides unique insights into the properties of heavy-quark hadronisation amid large partonic densities, where novel mechanisms beyond in-vacuum fragmentation may emerge. Examining heavy-flavour production across different collision systems and event multiplicities provides information about multi-parton interactions and the development of a strongly-interacting medium in high-multiplicity pp and p--Pb collisions. In Pb--Pb collisions, measurements of the nuclear modification factor ($R_{\rm AA}$) for charm and beauty hadrons provide a means to characterise the in-medium energy loss of heavy quarks in the quark-gluon plasma (QGP). In addition, measurements of the elliptic flow ($v_{2}$) for heavy quarks provide insights into their diffusion and their participation in the collective motion of the QGP. In this contribution, recent results of charm and beauty production measured with the ALICE detector are presented.
}

\FullConference{31st International Workshop on Deep Inelastic Scattering (DIS2024)\\
 8–12 April 2024\\
Grenoble, France\\}

\begin{document}
\maketitle

\section{Introduction}
The ALICE apparatus has been designed to study the quark-gluon plasma (QGP), a hot and dense medium in which quarks and gluons are deconfined. Open heavy-flavour hadrons, which contain one charm or beauty quark, are unique probes to investigate the properties of the QGP as heavy quarks are produced at the initial hard scatterings of a collision due to their large masses ($m_{\rm b}<m_{\rm c}<<\Lambda_{\rm QCD}$) and can therefore experience the entire evolution of the medium created in high energy heavy-ion collisions. Heavy-flavour measurements in Pb--Pb collisions allow us to explore the QGP properties via the medium effects on the heavy quarks traversing it, which result in a net energy loss of these quarks. In p--Pb collisions, the measurements help us to disentangle between initial and final state effects, and to test the interplay between soft and hard processes. The measurements in pp collisions serve as reference to measurement in p--Pb and Pb--Pb collisions, and allow us to test perturbative QCD (pQCD) calculations. Additionally, heavy-flavour hadrons are useful to reveal the details of heavy-quark fragmentation. In these proceedings, some of the recent results of heavy-flavour production measured by ALICE are presented.

\section{Results}

\begin{figure}[h!]
    \centering
    \includegraphics[height=0.4\textwidth]{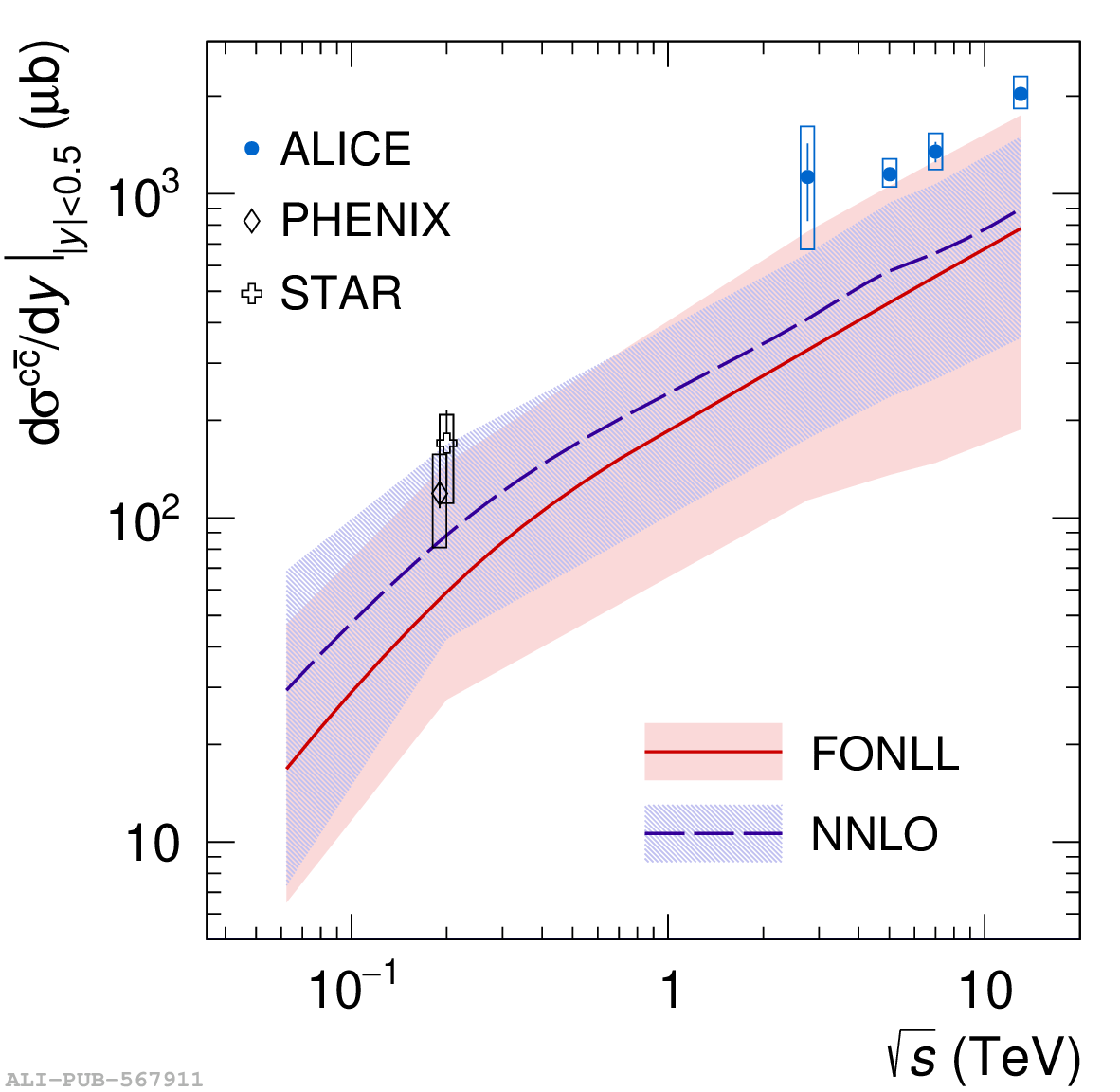}
    \includegraphics[height=0.4\textwidth]{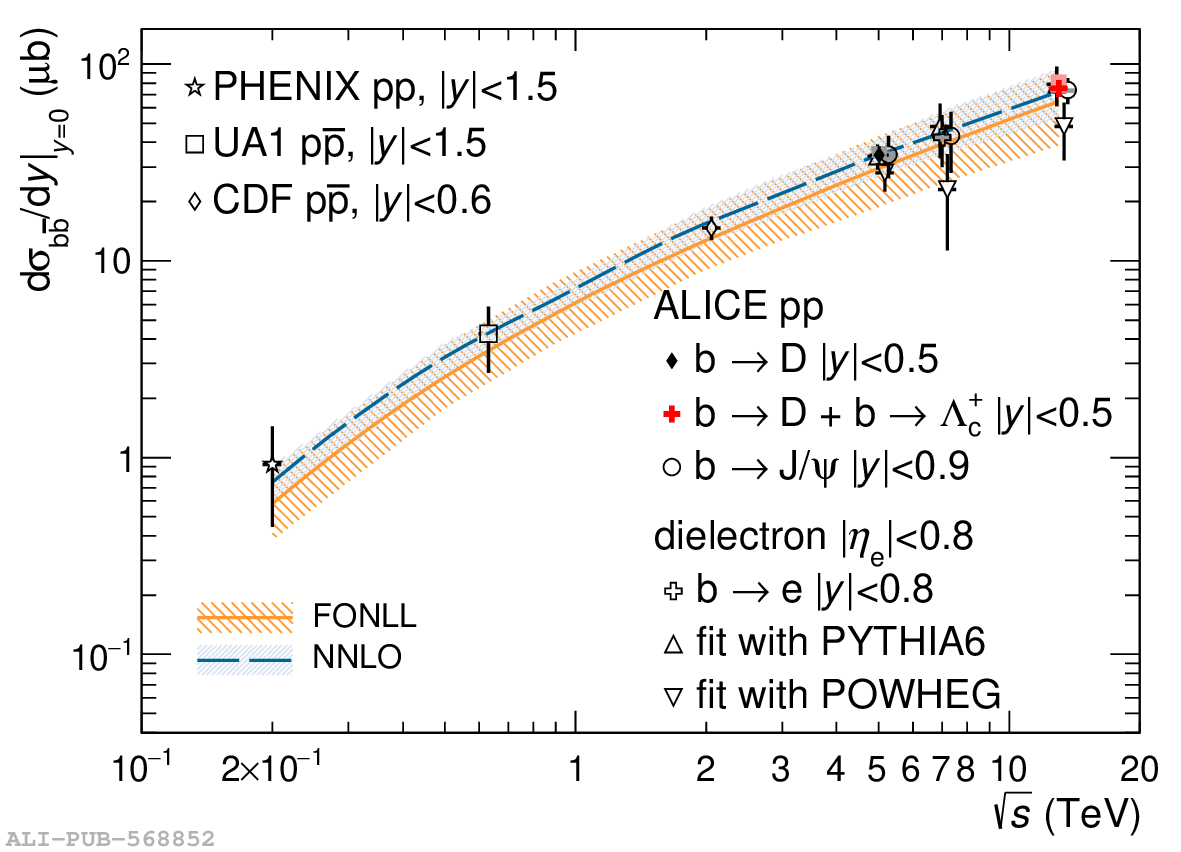}
    \caption{${\rm c}{\rm \bar{c}}$ (left~\cite{2023}) and ${\rm b}{\rm \bar{b}}$ (right~\cite{alicecollaboration2024measurement}) quark-pair production cross section per unit of rapidity at midrapidity in pp collisions as a function of $\sqrt{s}$. Both measurements are compared to predictions from FONLL~\cite{Cacciari_1998,Cacciari_2001,Cacciari_2012} and NNLO~\cite{denterria2016triple,d_Enterria_2017,Czakon_2013,Catani_2021} pQCD calculations.}
    \label{fig:CrossSection_vs_energy}
\end{figure}

Figure~\ref{fig:CrossSection_vs_energy} (left) shows the ${\rm c\Bar{c}}$ quark-pair production cross section at midrapidity ($|\eta|<0.5$) as a function of the centre-of-mass energy in pp collisions. The measurements are compared to FONLL~\cite{Cacciari_1998,Cacciari_2001,Cacciari_2012} and NNLO~\cite{denterria2016triple,d_Enterria_2017,Czakon_2013} pQCD calculations. The results at the LHC are systematically higher than the predictions, however, the measured cross sections are compatible with the theory uncertainty band within the current experimental precision. In Fig.~\ref{fig:CrossSection_vs_energy} (right), the ${\rm b\Bar{b}}$ quark-pair production cross section as a function of the center-of-mass energy in pp collisions at midrapidity is shown. The results are compared to the predictions from FONLL~\cite{Cacciari_1998,Cacciari_2001,Cacciari_2012} and NNLO~\cite{Catani_2021} calculations, and are compatible within the theoretical uncertainties. The NNLO calculations show smaller uncertainties than the FONLL calculations, and their central values are closer to the data due to the higher perturbative accuracy. These results provide constraints on the pQCD calculations as well as on the PDFs (Parton Distribution Functions).

\begin{figure}[h!]
    \centering
    \includegraphics[height=0.4\textwidth]{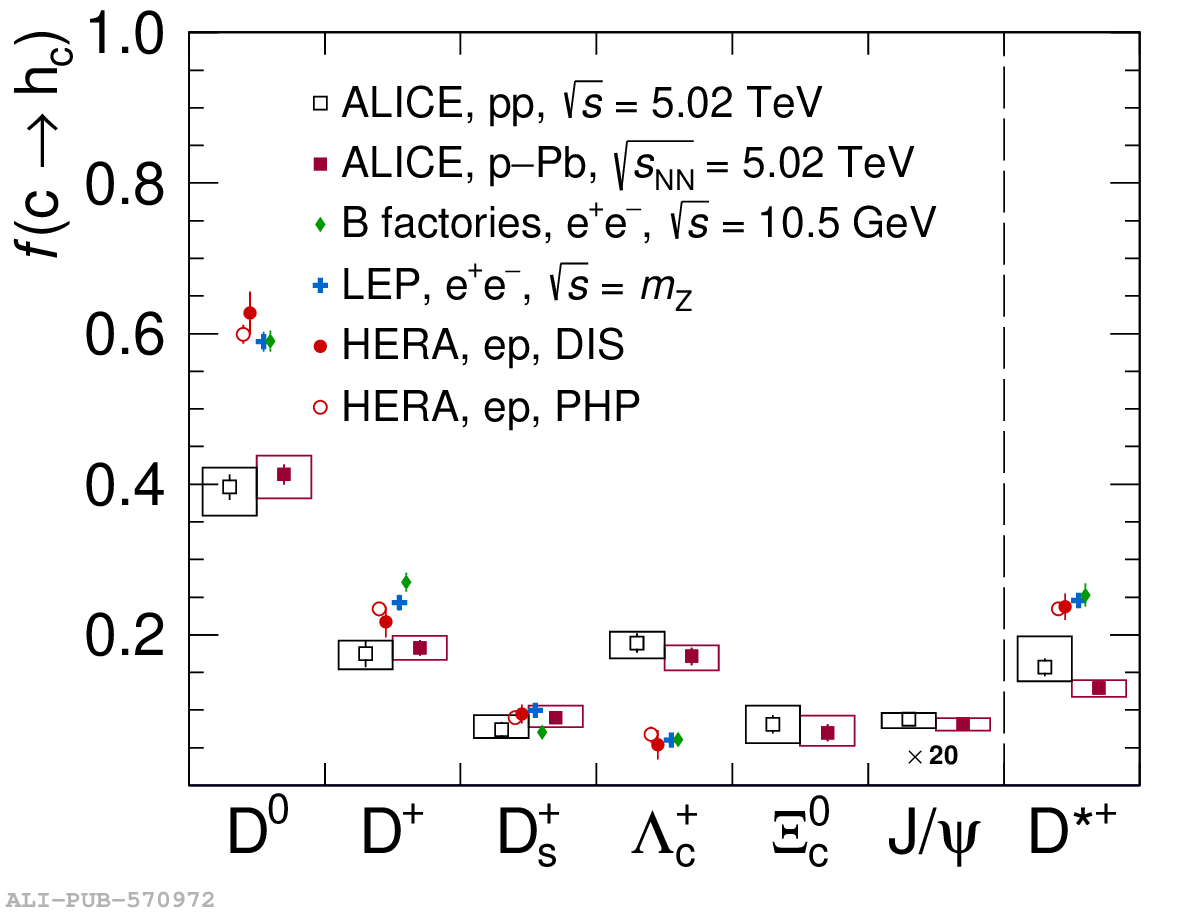}
    \includegraphics[height=0.4\textwidth]{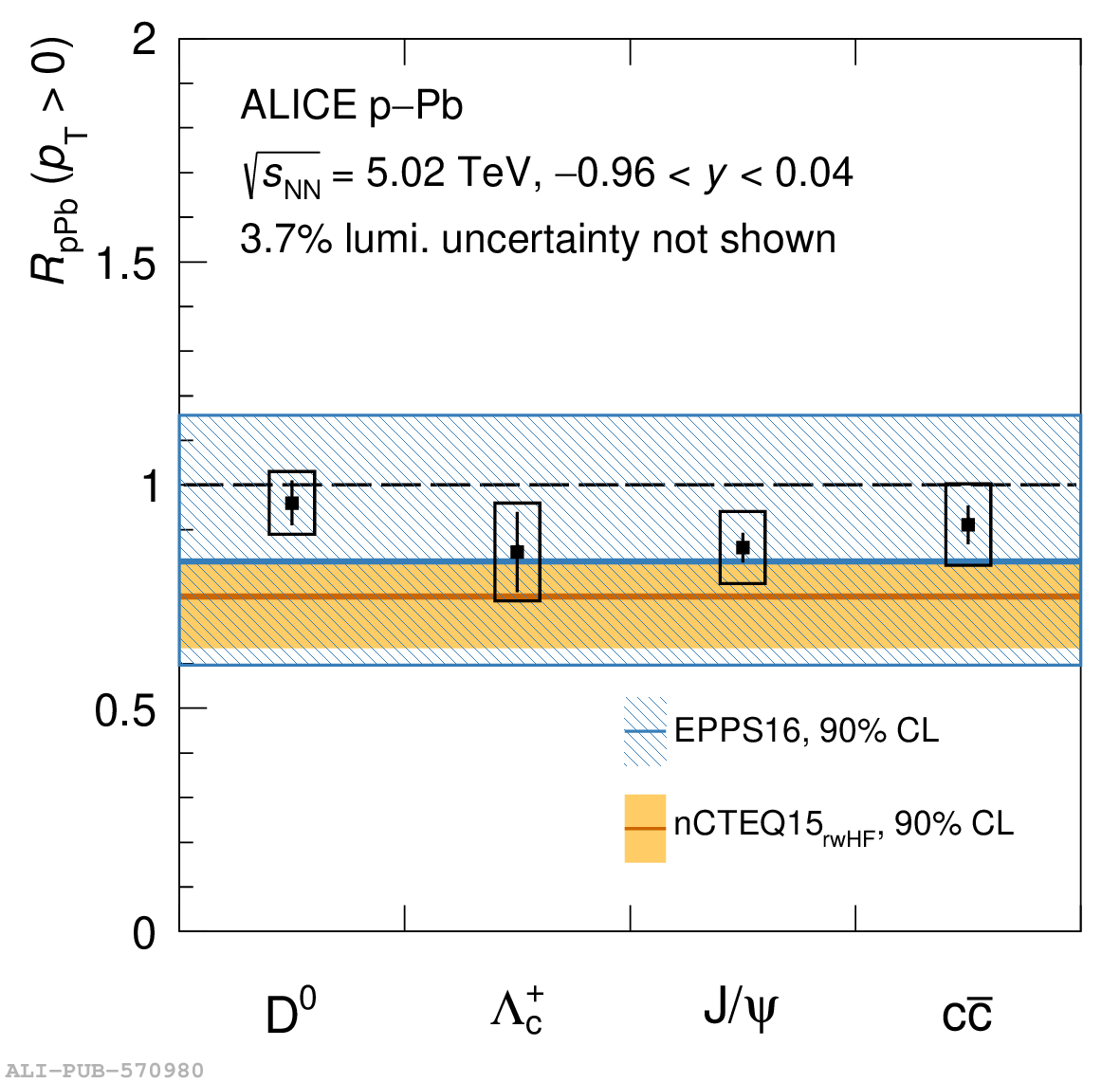}
    \caption{Left: The fragmentation fractions for charm hadrons in pp~\cite{charmpp5tev} and p--Pb~\cite{alicecollaboration2024charm} collisions at $\sqrt{s_{\rm NN}}=5.02\,{\rm TeV}$, compared to ones from e$^{+}$e$^{-}$ and ep collisions at lower energies~\cite{Lisovyi_2016}. Right: The $p_{\rm T}$-integrated nuclear modification factor $R_{\rm pPb}$ for ${\rm D}^{0}$, $\Lambda_{\rm c}^{+}$, ${\rm J}/\psi$, and ${\rm c\Bar{c}}$ pairs in p--Pb collisions at $\sqrt{s_{\rm NN}}=5.02\,{\rm TeV}$, compared to theoretical calculations using nCTEQ15$_{\rm rwHF}$~\cite{Kova_k_2016,Kusina_2018} and EPPS16~\cite{Eskola_2017} nPDF sets.}
    \label{fig:CrossSection_pPb}
\end{figure}
Figure~\ref{fig:CrossSection_pPb} (left) shows the fragmentation fractions for charm hadrons, i.e., the fraction of quarks fragmenting into a specific hadron among all measured ground-state charm-hadron species in p--Pb collisions at $\sqrt{s_{\rm NN}}=5.02\,{\rm TeV}$~\cite{alicecollaboration2024charm} compared to those measured in pp collisions at $\sqrt{s}=5.02\,{\rm TeV}$~\cite{charmpp5tev}. In spite of the larger system size and higher average charged-particle multiplicity, no significant modification of the hadronisation between pp and p--Pb collisions is observed. The measurements are also compared to the results from ${\rm e}^{+}{\rm e}^{-}$ and ${\rm e}{\rm p}$ collisions~\cite{Lisovyi_2016}. The prompt $\Lambda_{\rm c}^{+}$-baryon fragmentation fraction at the LHC is about three times larger than in ${\rm e}^{+}{\rm e}^{-}$ and ${\rm e}{\rm p}$ collisions, and the $\Xi_{\rm c}^{0,+}$ baryons account for about 10\% of the total charm hadron production at the LHC while these were assumed to be significantly smaller in ${\rm e}^{+}{\rm e}^{-}$ and ${\rm e}{\rm p}$ collisions. This enhancement of baryon production induces a corresponding deficit in the fragmentation fractions of D-mesons with respect to ${\rm e}^{+}{\rm e}^{-}$ and ${\rm e}{\rm p}$ collisions. The difference of the charm-hadron fragmentation fractions between pp, p--Pb and ${\rm e}^{+}{\rm e}^{-}$, ${\rm e}{\rm p}$ collisions indicates that the assumption of universal parton-to-hadron fragmentation is not valid.

The $p_{\rm T}$-integrated nuclear modification factors $R_{\rm pPb}$ for ${\rm D}^{0}$, $\Lambda_{\rm c}^{+}$, ${\rm J}/\psi$, and ${\rm c\bar{c}}$ in p--Pb collisions at $\sqrt{s_{\rm NN}}=5.02\,{\rm TeV}$ are shown in the right panel of Fig.~\ref{fig:CrossSection_pPb}. The $R_{\rm pPb}$ for ${\rm c\bar{c}}$ is consistent with unity within the measured uncertainties, which implies that the overall charm production rate in p--Pb collisions is consistent with that in pp collisions, indicating no sizeable nuclear modification effects. The $R_{\rm pPb}$ values for all charm hadrons are also consistent with unity within uncertainties, implying that the production rates of the individual charm hadrons are not strongly affected by nuclear effects. The experimental results are compared to the $p_{\rm T}$-integrated $R_{\rm pPb}$ calculated using both EPPS16~\cite{Eskola_2017} and nCTEQ15$_{\rm rwHF}$~\cite{Kova_k_2016,Kusina_2018} nPDF (nuclear Parton Distribution Function) sets, and are described by both of the calculations within uncertainties.

\begin{figure}[!h]
    \centering
    \includegraphics[height=0.3\textwidth]{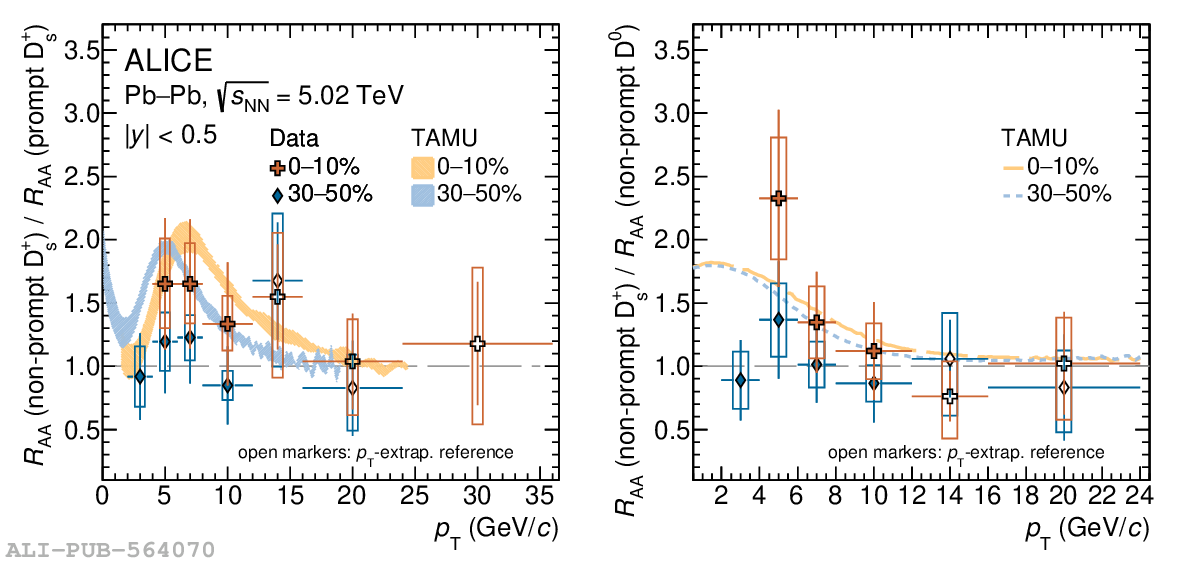}
    \includegraphics[height=0.3\textwidth]{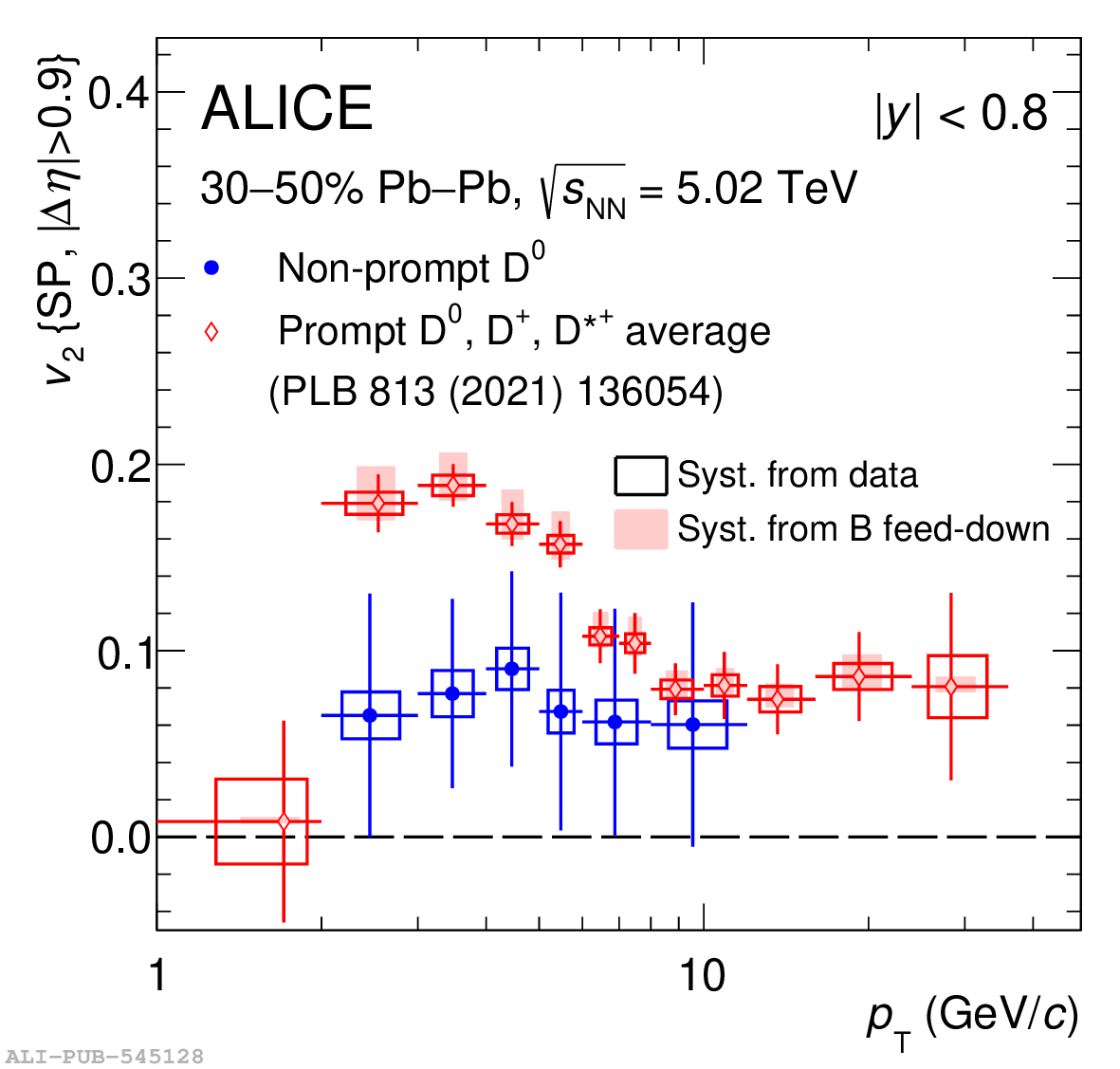}
    \caption{$R_{\rm AA}$ ratio of non-prompt ${\rm D}_{\rm s}^{+}$ to prompt ${\rm D}_{\rm s}^{+}$ (left) and non-prompt ${\rm D}^{0}$ (middle) for the 0--10\% and 30--50\% centrality intervals in Pb--Pb collisions at $\sqrt{s_{\rm NN}}=5.02\,{\rm TeV}$~\cite{PbPb_raa} compared to TAMU models~\cite{He_2014}. Right: Elliptic-flow coefficient $v_{2}$ of non-prompt ${\rm D}^{0}$ mesons and prompt non-strange D mesons in 30--50\% Pb--Pb collisions at $\sqrt{s_{\rm NN}}=5.02\,{\rm TeV}$~\cite{PbPb_flow}.}
    \label{fig:PbPb}
\end{figure}

In Fig.~\ref{fig:PbPb}, the ratios of the nuclear modification factor $R_{\rm AA}$ of non-prompt ${\rm D}_{\rm s}^{+}$ mesons to that of prompt ${\rm D}_{\rm s}^{+}$ mesons (left) and non-prompt ${\rm D}^{0}$ mesons (middle) in central (0--10\%) and semi-central (30--50\%) Pb--Pb collisions at $\sqrt{s_{\rm NN}}=5.02\,{\rm TeV}$~\cite{PbPb_raa} are shown. In the 0--10\% centrality class, the $R_{\rm AA}$ ratio of non-prompt ${\rm D}_{\rm s}^{+}$ to prompt ${\rm D}_{\rm s}^{+}$ is greater than unity in the $4<p_{\rm T}<12\,{\rm GeV}/c$ interval, suggesting a larger energy loss for charm quarks with respect to beauty quarks, while the ratio is consistent with unity in the 30--50\% centrality class. The measurements of the $R_{\rm AA}$ ratio of the non-prompt ${\rm D}_{\rm s}^{+}$ to non-prompt ${\rm D}^{0}$ suggest a hint of enhancement in the $4<p_{\rm T}<8\,{\rm GeV}/c$ interval for the 0--10\% centrality. The enhancement might be a consequence of the large abundance of strange quarks thermally produced in the QGP and the recombination-dominated hadronisation in this momentum range~\cite{Altmann:2024kwx}. The measurements are compared to the TAMU model predictions~\cite{He_2014}, which correctly describe the data within the experimental uncertainties.

The elliptic flow coefficient $v_{2}$ of non-prompt ${\rm D}^{0}$ mesons and the average $v_{2}$ of prompt ${\rm D}^{0}$, ${\rm D}^{+}$, and ${\rm D}^{*+}$ mesons in 30--50\% central Pb--Pb collisions at $\sqrt{s_{\rm NN}}=5.02\,{\rm TeV}$ are shown in Fig.~\ref{fig:PbPb}. The non-prompt ${\rm D}^{0}$ meson $v_{2}$ is larger than 0 with $2.7\sigma$ significance and no significant $p_{\rm T}$ dependence is observed. The non-prompt ${\rm D}^{0}$ meson $v_{2}$ is lower than that of prompt non-strange D mesons in $2<p_{\rm T}<8\,{\rm GeV}/c$ with a significance of $3.2\sigma$, indicating a weaker degree of participation in the collective motion of beauty quarks with respect to charm quarks.

\section{Summary}
The ALICE Collaboration has performed several heavy-flavour hadron measurements in pp, p--Pb, and Pb--Pb collisions at the LHC. In pp collisions, charm baryon enhancement is observed, resulting in larger charm fragmentation fractions for baryons with respect to ${\rm e}^{+}{\rm e}^{-}$ and ${\rm e}{\rm p}$ collisions. These results suggest that the hadronisation mechanisms are not universal and depend on collision systems. In p--Pb collisions, cold-nuclear matter effects do not significantly affect charm quark and charm hadron production. In Pb--Pb collisions, the $R_{\rm AA}$ of heavy-flavour hadrons provides a hint of the mass-dependent in-medium parton energy loss and hadronisation mechanism of heavy quarks. With Run3 data, heavy flavour measurements will be achieved with smaller uncertainties and extended transverse momentum reach, improving our understanding of heavy flavour production and hadronisation. 






\end{document}